\documentclass[aps,preprint,groupedaddress,showpacs]{revtex4}

\usepackage[dvips]{graphicx}
\usepackage{amsmath}
\usepackage{amssymb}
\usepackage{amsfonts}

\begin{document}

\date{\today}
\title{On the Suppression of Parametric Resonance and the Viability of Tachyonic Preheating after Multi-Field Inflation  }
\author{Diana Battefeld} 
\email{diana.battefeld(AT)helsinki.fi}
\affiliation{Helsinki Institute of Physics, P.O. Box 64, FIN-00014 Helsinki, Finland}
\affiliation{APC, UMR 7164, 10 rue Alice Domon et Leonie Duquet, 75205 Paris Cedex 13, France}
\author{Thorsten Battefeld}
\email{tbattefe(AT)princeton.edu}
\affiliation{Princeton University,
Department of Physics,
NJ 08544
}
\author{John T.~Giblin, Jr.}
\email{jgiblin(AT)bates.edu}
\affiliation{Department of Physics and Astronomy, Bates College, 44 Campus Ave, Lewiston, ME 04240}
\affiliation{The Perimeter Institute for Theoretical Physics, 31 Caroline St N, Waterloo, ON  N2L 2Y5, CANADA}

\begin{abstract}
We investigate the feasibility of explosive particle production via parametric resonance or tachyonic preheating in multi-field inflationary models by means of lattice simulations. We observe a strong suppression of resonances in the presence of four-leg interactions between the inflaton fields and a scalar matter field, leading to insufficient preheating when more than two inflatons couple to the same matter field. This suppression is caused by a dephasing of the inflatons that increases the effective mass of the matter field.

Including three-leg interactions leads to tachyonic preheating, which is not suppressed by an increase in the number of fields. If four-leg interactions are sub-dominant, we observe a slight enhancement of tachyonic preheating. Thus, in order for preheating after multi-field inflation to be efficient, one needs to ensure that three-leg interactions are present. If no tachyonic contributions exist, we expect the old theory of reheating to be applicable.

\end{abstract}
\maketitle
\newpage


\section{Introduction}

After inflation, the universe needs to heat up to temperatures above $1$MeV in order for primordial nucleosynthesis to commence. This temperature increase could occur via the perturbative decay of the inflaton(s) as described by the old theory of reheating \cite{Dolgov:1982th,Abbott:1982hn},  or by the rapid production of particles during a phase of preheating (see \cite{Bassett:2005xm,Kofman:2008zz} for reviews). The latter mechanism usually involves resonances or instabilities, with a resulting temperature well above $1$MeV, sometimes as high as $10^{15}$GeV, as opposed to the moderate temperatures accessible by means of the old theory of reheating \footnote{This perturbative decay could result in unwanted, long lived scalar particles,
 similar to the cosmological moduli problem \cite{Banks:1993en,de Carlos:1993jw}.}. High temperatures may be advantageous for scenarios of GUT baryogenesis (see i.e. \cite{Greene:1997ge}), but are usually in conflict with relic-bounds, one of which originates, for instance, from a prevention of gravitino-overproduction \cite{Ellis:1984eq,Bolz:2000fu}  ($T_{rh}\lesssim 10^9$GeV). For these reasons it is crucial to understand whether or not preheating occurs after inflation. 

Our current knowledge of preheating is primarily based on field theoretical models describing the inflatons' decay by non-perturbative effects, such as parametric resonance \cite{Traschen:1990sw,Felder:2000hj,Kofman:1997yn} or tachyonic preheating \cite{Dufaux:2006ee} among others (see \cite{Bassett:2005xm,Kofman:2008zz} for reviews). This approach should be seen as a preliminary step in the absence of a better understanding of the degrees of freedom that drive and end inflation. For instance, it may be that stringy effects are crucial, such as in brane inflation models where inflation ends once branes annihilate \cite{Kofman:2005yz}. The hope is that an effective field theoretical approach emerges, once we have a better understanding of the microphysical nature of inflation.  In the meantime, it is desirable to have a thorough understanding of the conditions under which the energy transfer from the inflationary sector to preheat matter fields is efficient. Further, the preliminary theory of preheating has been successfully applied in other areas, for instance in the notion of moduli trapping on the landscape \cite{Kofman:2004yc,Watson:2004aq,Greene:2007sa}, a possible solution \cite{Shuhmaher:2007pv} to the moduli-induced gravitino problem \footnote{This problem arises in the heavy moduli scenario \cite{Endo:2006zj,Nakamura:2006uc}, a solution to the cosmological moduli-problem \cite{Coughlan:1983ci,Ellis:1986zt,de Carlos:1993jw,Banks:1993en}.}, or as an additional source of gravitational waves \cite{Khlebnikov:1997di,GarciaBellido:1998gm,Easther:2006vd,Easther:2007vj,Dufaux:2007pt,Easther:2006gt,Dufaux:2008dn,Price:2008hq}.
In light of these applications, it is necessary to investigate the viability of the proposed mechanisms, particularly as, in recent years, new developments in string cosmology (see \cite{McAllister:2007bg,Cline:2006hu,Burgess:2007pz,Kallosh:2007ig} for recent reviews) have led to the emergence of several multi-field inflationary models; a few examples are inflation on the landscape \cite{Battefeld:2008qg,Tye:2008ef}, inflation driven by axions i.e.~$\mathcal{N}$-flation \cite{Dimopoulos:2005ac,Easther:2005zr} (see also \cite{Misra:2007cq,Misra:2008tx}), inflation driven by tachyons \cite{Piao:2002vf,Majumdar:2003kd} or by multiple M5-branes \cite{Becker:2005sg,Krause:2007jr,Ashoorioon:2006wc,Ashoorioon:2008qr} among others (see \cite{Wands:2007bd} for a review).  
The majority of these models are realizations of assisted inflation \cite{Liddle:1998jc,Malik:1998gy,Kanti:1999vt,Kanti:1999ie,Calcagni:2007sb}, where many fields are dynamical. An advantage of these models is the possible alleviation of the $\eta$-problem, since any given field does not need to traverse a large distance in field space, given that many  fields ($\mathcal{N}\sim 10^3$) assist each other in driving the inflationary phase. 

Unfortunately, little is known about the actual process of reheating standard model (SM) degrees of freedom in almost all inflationary scenarios, and there is a danger of primarily heating hidden sectors, especially if many fields contribute to inflation (see \cite{Green:2007gs} for a case study within $\mathcal{N}$-flation). If inflation would be driven with the field content of the MSSM \cite{Allahverdi:2006iq,Allahverdi:2006cx,Allahverdi:2006we,Allahverdi:2008bt} the standard model can be properly reheated \cite{Allahverdi:2006we}, for instance via instant preheating \cite{Felder:1998vq,Allahverdi:2006we}, since the couplings of these fields are known. However, it is challenging to achieve the needed sixty e-folds of inflation within this scenario; as a result,  a prior phase of high-scale inflation is usually assumed, commonly terminated by a phase of preheating.
A related possibility to reheat primarily standard model particles, which does not involve low scale inflation, consists of a second phase of reheating, caused by the decay of a long lived, light particle species such as the saxion (see i.e. \cite{Banks:2002sd,Heckman:2008jy}\footnote{One can also reheat via MSSM flat directions \cite{Enqvist:2003qc}; see also \cite{Acharya:2008bk}.}), which dilutes any unwanted relics and overabundance of hidden sector particles produced during preheating. 

However, even a field theoretical description of preheating is by no means complete; we lack an understanding to which fields the inflatons couple, what type of couplings are present, and how big the coupling constants are.
Non-perturbative mechanisms are common, given a certain amount of fine tuning, but they have mainly been studied in simple setups involving one \cite{Traschen:1990sw,Dolgov:1989us,Kofman:1997yn,Greene:1997fu} or two \cite{Bassett:1998yd,Bassett:2005xm,Bassett:1997gb} inflatons coupled to a single matter field, or one inflaton coupled to several matter fields \cite{Anderson:2008dg} (see however the recent case study of preheating after $\mathcal{N}$-flation with $\mathcal{N}\sim 100$ fields \cite{Battefeld:2008bu,Battefeld:2008rd}). Preheating is complicated by the importance of backreaction after the first few bursts of particle production which affects resonances; these effects can only be recovered using lattice simulations \cite{Prokopec:1996rr,Felder:2000hq,Frolov:2008hy,Khlebnikov:1996zt,Khlebnikov:1996mc}, which have only become feasible for more complicated scenarios in recent years. Further, even if the inflatons decay via this route, the universe is still in a non-thermal state.  The era of thermalization is involved and again requires lattice simulations \cite{Felder:2000hr,Podolsky:2005bw}.

In this article, we investigate the early stages of preheating in multi-field inflationary models via a lattice simulation, incorporating backreaction and up to five inflatons. We use an implementation of {\sc LatticeEasy} \cite{Felder:2000hq} ({\sc Defrost}, an alternative code developed by A.~Frolov, can be found in \cite{Frolov:2008hy}). Our aim is to investigate conditions under which parametric resonance (quadratic couplings between matter and inflaton fields, that is four-leg interactions) and tachyonic preheating (Yukawa couplings, that is three-leg interactions) are efficient. By efficient we mean that the majority of the energy in the inflaton fields is transferred to preheat matter fields within a few hundred oscillations. Our primary concern is the alteration of resonance effects or instabilities if more than one inflaton field is coupled to the same matter field. For instance, in the presence of a few fields, the expectation is that resonances are enhanced, since stability bands are generically destroyed, leading to instabilities of almost all Fourier modes of the matter field; this is sometimes referred to as Cantor Preheating \cite{Bassett:1997gb} and has been studied numerically in a two-field model (without backreaction) in \cite{Bassett:1998yd}. However, in the presence of many fields, inflatons generically run out of phase, which in turn increases the effective mass of the matter field so that efficient preheating becomes impossible \cite{Battefeld:2008bu,Battefeld:2008rd}. Our first goal is to understand quantitatively the competition of these two opposing effects as we increase the number of fields one by one. 
Secondly, we examine tachyonic preheating (three-leg interactions), which is not expected to be hindered  by dephasing effects \cite{Braden}. In the presence of both, three- and four-leg interactions, it is unclear if an increase of the matter fields effective mass leads to a suppression of instabilities.

We find that resonances are strongly suppressed in the presence of four-leg interactions if more than two inflatons couple to the same preheat matter field; further, even in the presence of only two inflatons, resonances are not enhanced, as indicated by Bassett et.al. in \cite{Bassett:1998yd}, but slightly suppressed. Thus, Cantor preheating is less efficient than previously anticipated. Turning our attention to Yukawa couplings, we find that particle production remains efficient. There is no suppression if the number of inflatons is increased, and if four-leg interactions are suppressed, tachyonic preheating is sightly enhanced.

The outline of this paper is as follows: after introducing a generic but simple multi-field inflationary model in Sec.~\ref{sec:preheating}, we provide the results of lattice simulations for standard parametric resonance models based on four-leg interactions in Sec.~\ref{caseA}, mixed three- and four-leg interactions in Sec.\ref{caseB} and tachyonic preheating (three-leg interactions dominate) in Sec.~\ref{caseC}. We conclude in Sec.\ref{sec:discussion}. Details about our implementation of {\sc LatticeEasy}  can be found in Appendix \ref{appendix:code}. Throughout this article we set the Planck mass equal to one, $m_{pl}^2=1/G\equiv 1$.

\section{Preheating \label{sec:preheating}}
Consider several inflaton fields $\varphi_i$, $i=1\dots \mathcal{N}$, with canonical kinetic terms. After inflation, each inflaton approaches its respective minimum in the potential. Since we are interested in preheating, we expand the potential around the minima so that, after redefining the fields such that the global minimum lies at $\varphi_i=0$, the total potential becomes $W=\sum_i V_i$ with $V_i=m_i^2\varphi_i^2/2$; generically $m_i\neq m_j$ for $i\neq j$, so that the fields oscillate with different frequencies. In simple models of inflation the slope during preheating is related to the slope during inflation. These models are also the most predictive ones \cite{Boyle:2008ri,Boyle:2005ug,Bird:2008cp}, since more ad hoc features in the potential allow the fitting of more or less any exotic data. For instance, if inflation is driven by a single field with a quadratic potential (the simplest model imaginable), its mass is determined by the COBE normalization $P_{\zeta}=(2.41\pm 0.11) \times 10^{-9}$ \cite{Komatsu:2008hk}; for more than one field with simple quadratic potentials the effective single field still needs to satisfy the COBE normalization, but the actual spread of masses is not significantly constrained  \footnote{In some concrete scenarios the mass distribution is known (i.e.~in $\mathcal{N}$-flation \cite{Easther:2005zr}), and should be used during preheating (see i.e.~\cite{Battefeld:2008bu}).}. Heavy fields become irrelevant during inflation, since they quickly roll down the potential. As a consequence, we expect a narrow spread of masses for fields relevant for preheating, at least in simple models (this is indeed the case in $\mathcal{N}$-flation \cite{Battefeld:2008bu}). Based on this reasoning, we distribute $m_i^2$ equidistantly over the interval $0.5 \times 10^{-12}\dots 1.5 \times 10^{-12}$, i.e. $m_1^2=10^{-12}$ for $\mathcal{N}=1$, $m_1^2=0.5\times 10^{-12}$ and $m_2^2=1.5\times 10^{-12}$ for $\mathcal{N}=2$, $m_1^2=0.5\times 10^{-12},m_2^2=10^{-12}$ and $m_3^2=1.5\times 10^{-12}$ for $\mathcal{N}=3$ etc. Note that we ordered the masses: $m_i<m_j$ if $i<j$. We further define the average square mass as $m^2\equiv 10^{-12}$ and a dimensionless time via $\tau \equiv t m$ (see Appendix \ref{appendix:code}); the latter one is used in all plots. 

In realistic scenarios, it may be that several fields decay during inflation \cite{Battefeld:2008py,Battefeld:2008ur,Battefeld:2008qg} or assist inflation only during early stages \cite{Battefeld:2008bu}. Such fields do not make a considerable contribution  to preheating and are therefore excluded. The fields we consider should have comparable energies when preheating commences. For simplicity, we impose equal energy initial conditions $m_i^2\varphi_i^2(0)=m_j^2\varphi_j^2(0)$ and set the total potential energy equal to $\sum_i m_i^2\varphi_i^2(0)/2\equiv (10^{-6})^2\times(0.193)^2/2\approx 1.8\times 10^{-14}$ at the onset of preheating, the usual value for a single inflaton field with a quadratic potential \cite{Kofman:1997yn}. These initial conditions simplify comparisons of setups with a varying number of fields.

To model preheating, we couple all inflaton fields to the same scalar matter field $\chi$. In a realistic scenario we expect a more complicated matter sector, but focusing on only one matter field should provide an instructive toy model. We allow for Yukawa type interactions $\sigma_i\varphi_i\chi^2/2$ and quadratic interactions $g_i\varphi_i^2\chi^2/2$, since both are generically present \footnote{Note that three-leg interactions are generated if $\varphi_i=0 \, \forall i$  does not coincide with the location where $\chi$ becomes lightest according to four-leg interactions.}; however, we ignore terms higher order in $\varphi_i$ as well as cross coupling terms proportional to $\varphi_i\varphi_j$ for $i\neq j$ (the former ones are sub-leading in the Taylor expansion and the latter ones are usually suppressed for multi-field inflationary models of interest, such as assisted inflation \cite{Liddle:1998jc}). During preheating, only light degrees of freedom are produced \cite{Prokopec:1996rr,Zlatev:1997vd}, hence, we neglect the necessarily small bare mass of $\chi$. However, to guarantee that the potential is bounded from below for large $\chi$ and to increase the stability of our code, we include a small self interaction term $\lambda\chi^4/4$; this term prevents an unphysical runaway behavior of $\chi$ in our numerical integration without influencing preheating significantly. Because we expect coupling constants between the inflatons and $\chi$ to be comparable \footnote{Without a compelling theoretical reason, vastly different coupling constants constitute fine tuning.}, we take $g_i\equiv g$ and $\sigma_i\equiv \sigma$. All in all, the potential we consider during preheating reads
\begin{eqnarray}
W=\sum_{i=1}^{\mathcal{N}}\left(\frac{m_i^2}{2}\varphi_i^2+\frac{\sigma}{2}\varphi_i\chi^2+\frac{g}{2}\varphi_i^2\chi^2\right)+\frac{\lambda}{4}\chi^4\,.
\end{eqnarray}
The equations of motion are the Friedmann equations which we combine to
\begin{eqnarray}
\ddot{a}+2\frac{\dot{a}^2}{a}-\frac{8\pi}{a}\left(\frac{1}{3}\sum_{i=1}^{\mathcal{N}}\left|\nabla \varphi_i\right|^2+\frac{1}{3}\left|\nabla \chi\right|^2+a W\right)=0\,,
\end{eqnarray}
as well as the Klein-Gordon equations
\begin{eqnarray}
\ddot{\varphi}_i+3\frac{\dot{a}}{a}\dot{\varphi}_i-\frac{1}{a^2}\nabla^2 \varphi_i+\frac{\partial W}{\partial \varphi_i}=0\,,\\
\ddot{\chi}+3\frac{\dot{a}}{a}\dot{\chi}-\frac{1}{a^2}\nabla^2 \chi+\frac{\partial W}{\partial \chi}=0\,.
\end{eqnarray}
We integrate the above set numerically using  the {\sc LatticeEasy}  code \cite{Felder:2000hq} which uses a staggered leapfrog integrator. The computation is performed in real space on a $n^3=128^3$ lattice, with a comoving box size of $L=5/m$, while the initial conditions (including small initial inhomogeneities) are prescribed in Appendix \ref{appendix:code} (see also \cite{Felder:2000hq}). This means we can cover wave-numbers in the interval $2\pi / L < k < 2 \pi \sqrt{3}n/(2L)$. 

In the next sections, we discuss the results of various choices of  $\mathcal{N}$ and the coupling constants $\sigma$ and $g$. For more details on our implementation of {\sc LatticeEasy}  see Appendix \ref{appendix:code}.

\begin{table}
\begin{tabular*}{9.5cm}{@{\extracolsep{\fill}}|c|c|c|c|}
\hline
Case & $g\times (0.193)^2/m^2$& $\sigma \times 0.193/m^2$ & $\lambda\times (0.193)^2/m^2$  \\
\hline\hline
A& $10^4$ & $0$ & $5\times 10^3$  \\
B& $10^4$& $100$ & $5\times 10^3$  \\
C& $100$& $100$ & $10^4$  \\
\hline
\end{tabular*}
\caption{The coupling constants for cases A (four-leg interactions, Sec.~\ref{caseA}), B (three- and four-leg interactions, Sec.~\ref{caseB}) and C (three-leg interactions and minor four-leg interactions (added for stability), Sec.~\ref{caseC}). $\lambda\neq 0$ suppresses an unphysical runaway behavior of $\chi$, but does not interfere with preheating; here $m=10^{-6}$ and the Planck mass is set to one. For ease of comparison we chose values identical to those in \cite{Dufaux:2006ee}, where tachyonic preheating for $\mathcal{N}=1$ is discussed (note that instead of $g^2$ as in \cite{Dufaux:2006ee} we use $g$ in the potential).  }
\label{t:parameters}
\end{table}

\subsection{Parametric Resonance ($\sigma=0$, $g\neq 0$): Suppression due to Dephasing \label{caseA}}
\begin{figure}[tb]
\includegraphics[width=\textwidth,angle=0]{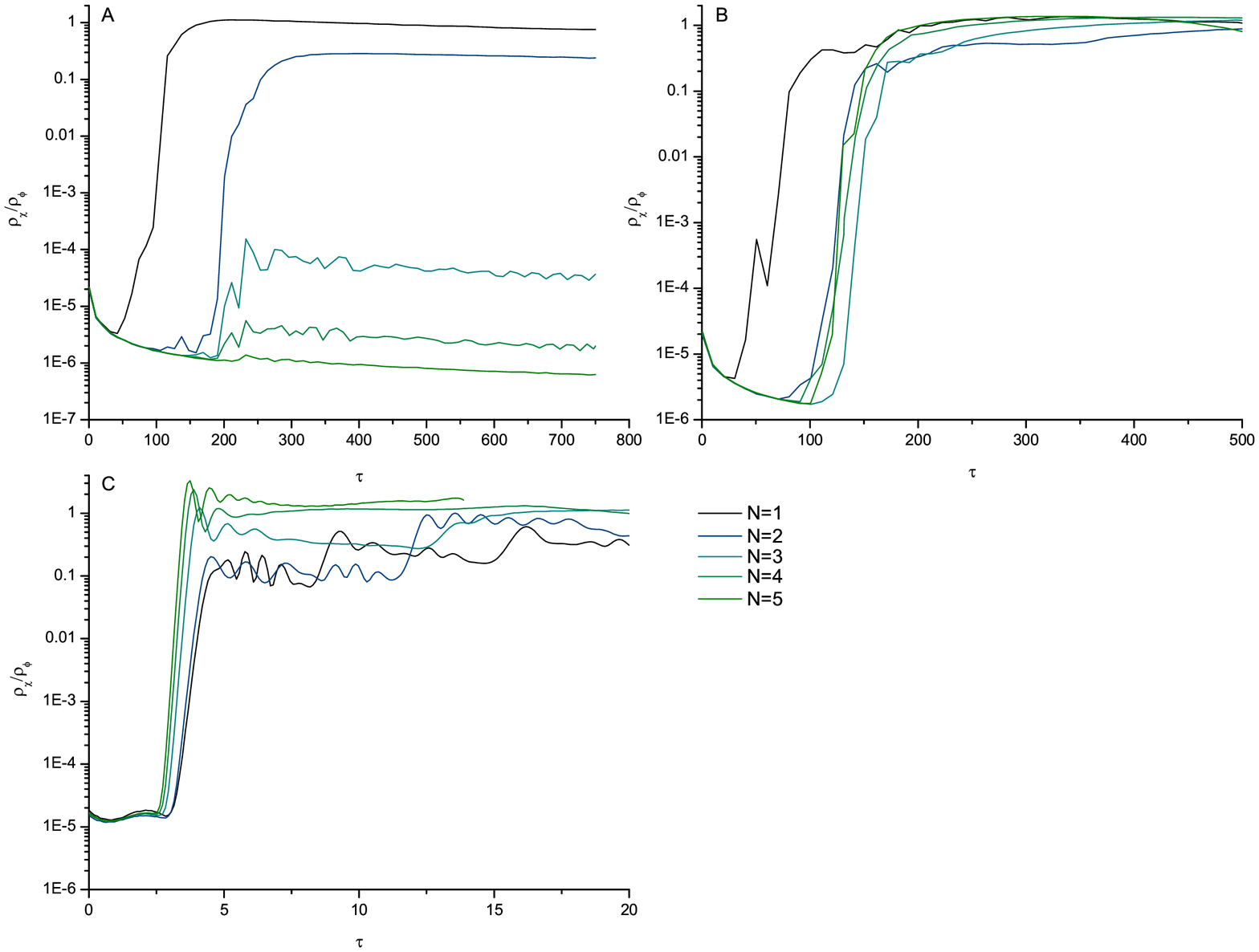}
   \caption{\label{pic:energy_ratio_A_B_C_N=128}
We plot the ratio $\rho_\chi/\rho_\varphi$ of the total matter energy density (kinetic, potential and gradient) to the total energy density in the inflatons $\rho_\varphi=\sum_i\rho_{\varphi_i}$ (note $\chi^2\sum_i(g\varphi_i^2+\sigma \varphi_i)/2$ is not included in either $\rho_\varphi$ or $\rho_\chi$) for cases A, B and C  ($n^3=128^3$-lattice, $L=5/m$, for coupling constants see Table \ref{t:parameters}). Preheating is successful if this ratio becomes of order one; a turbulent phase and an era of thermalization follow (not covered here). Before $\rho_\chi/\rho_\varphi\sim 1$, the fields fragment and backreaction becomes important, see Fig.~\ref{pic:variance_A} and \ref{pic:variance_B_C} for the variance of the fields. Panel A: Parametric resonance, and thus preheating, is strongly suppressed for $\mathcal{N}\geq 3$. Panel B: a comparison with panel A reveals that resonances are enhanced due to tachyonic preheating, which is caused by $\sigma\chi^2\sum_i \varphi_i/2$; an increase in the number of fields has no noticeable effect, except for a slight delay of particle production. Panel C: tachyonic preheating is slightly enhanced if the number of fields is increased.}
\end{figure}

\begin{figure}[tb]
\includegraphics[width=\textwidth,angle=0]{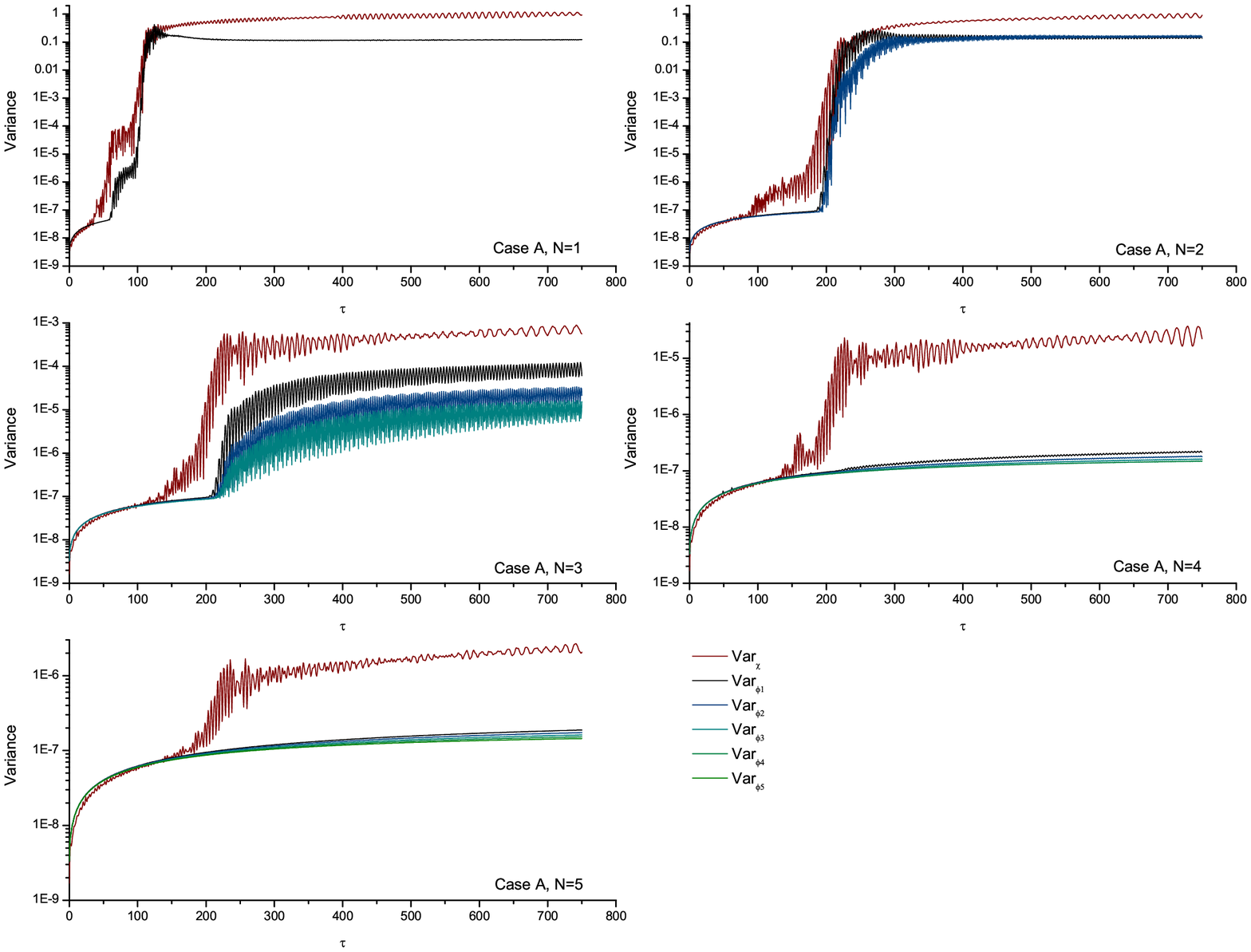}
  \caption{\label{pic:variance_A} The variance (\ref{variance}) of the fields for case A ($\sigma=0$, $g\neq 0$, $n^3=128^3$-lattice, $L=5/m$), increasing the number of fields from $\mathcal{N}=1$ to $5$. Fields fragment and backreaction becomes important once the variance is of order one. For $\mathcal{N}\geq 3$ the fields remain essentially homogeneous, due to the suppression of resonances caused by dephasing fields; preheating is inefficient in these cases.}
\end{figure}

In the absence of three-leg interactions and considering a single inflaton field  ($\mathcal{N}=1$), we observe the well known amplification of $\chi$ due to parametric resonance \cite{Dolgov:1989us,Dolgov:1989us,Kofman:1997yn} (Fig.~\ref{pic:energy_ratio_A_B_C_N=128} A and \ref{pic:variance_A} A, $\mathcal{N}=1$). To be concrete, if backreaction and expansion effects are ignored, the equation of motion for the Fourier modes of the matter field $\chi_k$ turns into the Mathieu equation
\begin{eqnarray}
\chi_k^{\prime\prime}+\left(A_k-2q\cos(2\tau)\right)\chi_k=0\,,
\end{eqnarray}
where a prime denotes a derivative with respect to $\tau=mt$, $A_k=k^2/m^2+2q$, $q=g\Phi^2/(4m^2)$ and $\Phi$ is the initial amplitude of the inflaton field oscillations $\varphi=\Phi \cos(\tau) $; the Mathieu equation exhibits
well known stability and instability bands \cite{Kofman:1997yn,McLachlan}, resulting in narrow or broad resonance \cite{Kofman:1997yn}. If a mode lies within an instability band, its amplitude grows exponentially $\propto e^{\mu_k}$, where $\mu_k>0$ is the Floquet index. Including the expansion of the universe causes modes to shift through bands, leading to stochastic resonance \cite{Kofman:1997yn}. Here, one can still derive approximate analytic expressions describing the growth of the matter field \cite{Kofman:1997yn}, given that only one inflaton field is involved. However, in all cases backreaction becomes important soon after the first spurts of particle production, causing the inflaton field to fragment and leading to an extended turbulent regime \cite{Khlebnikov:1996zt,Khlebnikov:1996mc}, followed by an even longer phase of thermalization \cite{Podolsky:2005bw,Felder:2000hr}. We checked that we recover these known results in our simulations (see for instance Fig.~\ref{pic:variance_A} for the variance (\ref{variance}) in the $\mathcal{N}=1$ case).

In the presence of two fields, Bassett argued that resonances could be enhanced: based on spectral theory \cite{JMoser, lachlan, gihman, Pastur, Avron, JAvron, JAvronBSimon}, he showed that the stability bands generically dissolve into a nowhere dense set, similar to a Cantor set, if the ratio of the inflatons' oscillation frequencies is not a simple fraction; then almost all modes should be amplified \cite{Bassett:1997gb}. This is sometimes referred to as Cantor preheating. However, no analytic expressions are known for the magnitude of the generalized Floquet index $\mu_k$. A numerical study followed in \cite{Bassett:1998yd}, where one peculiar two field model was investigated (backreaction was neglected) and indeed a slight enhancement was found. However, in the presence of many inflaton fields it was recently shown in \cite{Battefeld:2008bu} that resonances are generally suppressed: if inflatons dephase, which occurs quickly in most models even if the fields start to oscillate in unison, an increase of the matter field's effective mass results, because $m_{\chi (eff)}^{2}$ contains a term proportional to $\propto\sum_i\varphi_i^2$. Since the matter field turns heavy and the oscillations of the matter field's effective mass are smeared out, it becomes more difficult to produce $\chi$ particles. As a consequence, resonances are suppressed.
This suppression is strong for large $\mathcal{N}$, as seen in the case of $\mathcal{N}$-flation where $\mathcal{N}\sim 100$ \cite{Battefeld:2008bu}; this numerical result was confirmed in \cite{Braden}.
In \cite{Battefeld:2008bu} preliminary results were reported, indicating that fewer fields would still allow for resonances, which appeared to be weaker than in the single field case, in contrast to the findings of \cite{Bassett:1998yd}. 

Here, we provide a complete numerical simulation, including the expansion of the universe and backreaction effects, to settle this issue. It is evident from Fig.~\ref{pic:energy_ratio_A_B_C_N=128} A that resonances are indeed suppressed, even in the presence of only two fields. Thus, the initial hope that particle production would be enhanced in Cantor preheating is not true in general. The slight enhancement found in \cite{Bassett:1998yd} might have been caused by the chosen coupling constants in conjunction with ignoring backreaction. Further, for as few as three fields, parametric resonance is heavily suppressed and disappears entirely for as few as five fields over the time-scale of our simulations (we cover over one hundred oscillations of the inflaton fields). 

The onset of parametric resonance shifts to later times if $\mathcal{N}$ increases from one to two, but stays about the same if $\mathcal{N}$ is further increased, Fig.~\ref{pic:energy_ratio_A_B_C_N=128} A. This effect could be caused by our chosen initial conditions, which do not coincide precisely with the breakdown of slow roll if more than one inflaton is involved. We cannot exclude particle production on longer time scales, but since the universe is expanding all along, causing the energy in the oscillating inflaton fields to redshift, it becomes more difficult, and at some point impossible, to start preheating. As a result, the universe remains dominated by the oscillating inflaton fields until they decay perturbatively (i.e.~mediated by gravity).

The variance (\ref{variance}) of the matter and inflaton fields is plotted in Fig.~\ref{pic:variance_A}. A variance of order one indicates that a field is fragmented. We see that fields quickly become inhomogeneous, and backreaction becomes important, once explosive particle production occurs. For $\mathcal{N}\geq 3$ no fragmentation of the fields is evident, simply because particle production is absent in the first place.

We conclude that the well known parametric resonance models based on four-leg interactions ($g\varphi^2\chi^2/2$) are unlikely to work without fine tuning if more than two inflaton fields contribute to preheating. This conclusion is in line with the analytic arguments given in \cite{Battefeld:2008bu,Battefeld:2008rd}, but could be evaded by coupling each inflaton field to its own preheat matter field. However, such a construction seems somewhat artificial and ill motivated to us.

In the absence of preheating the old theory of reheating \cite{Dolgov:1982th,Abbott:1982hn} (perturbative decay over many oscillations) still applies, as proposed in \cite{Battefeld:2008bu}. However, a delayed decay of the inflatons might pose problems if a field survives long enough to interfere with the thermal history of the universe. This problem was pointed out in \cite{Braden} within $\mathcal{N}$-flation, and tachyonic preheating \cite{Dufaux:2006ee} was proposed as an alternative; this type of preheating should be less affected by dephasing, since the matter field's effective mass contains a term proportional to $\propto \sum_i\varphi_i$, which is still negative half of the time. We investigate this interesting possibility in the next sections. Our main concern is that the incorporation of a four-leg interaction ($g\neq 0$) in addition to a three-leg interaction ($\sigma\neq 0$) could again spoil preheating. 

We would like to point out that a suppression of preheating might be desirable: the reheating temperature after preheating is generally high enough to violate bounds originating from preventing gravitinos from over-closing the universe or spoiling the success of primordial nucleosynthesis \footnote{Often the gravitino is the lightest SUSY particle with a long lifetime, providing a candidate for dark matter. However, its decay and the resulting shower of energetic photons can cause the disintegration of light elements whose abundances, as predicted from nucleosynthesis, are in excellent agreement with observations (except Lithium, which is off by a factor of 2 to 5).}. The maximum reheating temperature is usually taken to be $T_{rh}\lesssim 10^{9} GeV$ \cite{Ellis:1984eq,Bolz:2000fu} (see i.e. \cite{Heckman:2008jy} for a pedagogical review). This problem can be resolved via a second phase of reheating: if a long lived light scalar field dominates the energy density of the universe after the decay of the inflaton(s), i.e. because it redshifts like matter due to oscillations in a quadratic potential, it causes a second phase of reheating once it decays; the dilution of any particles present prior to this second phase of reheating alleviates the strict bounds on the reheating temperature \cite{Banks:2002sd}. A concrete realization of this scenario is the decay of the saxion within the framework of F-theory \cite{Heckman:2008jy} (see also \cite{Acharya:2008bk} for related work withing G2-MSSM models). Given a multi-field inflationary scenario, the role of this late decaying field may be played by one of the inflatons that did not decay successfully during preheating.

\subsection{Generic Preheating Case ($\sigma\neq 0$, $g\neq 0$): no Suppression \label{caseB}}

\begin{figure}[tb]
\includegraphics[width=\textwidth,angle=0]{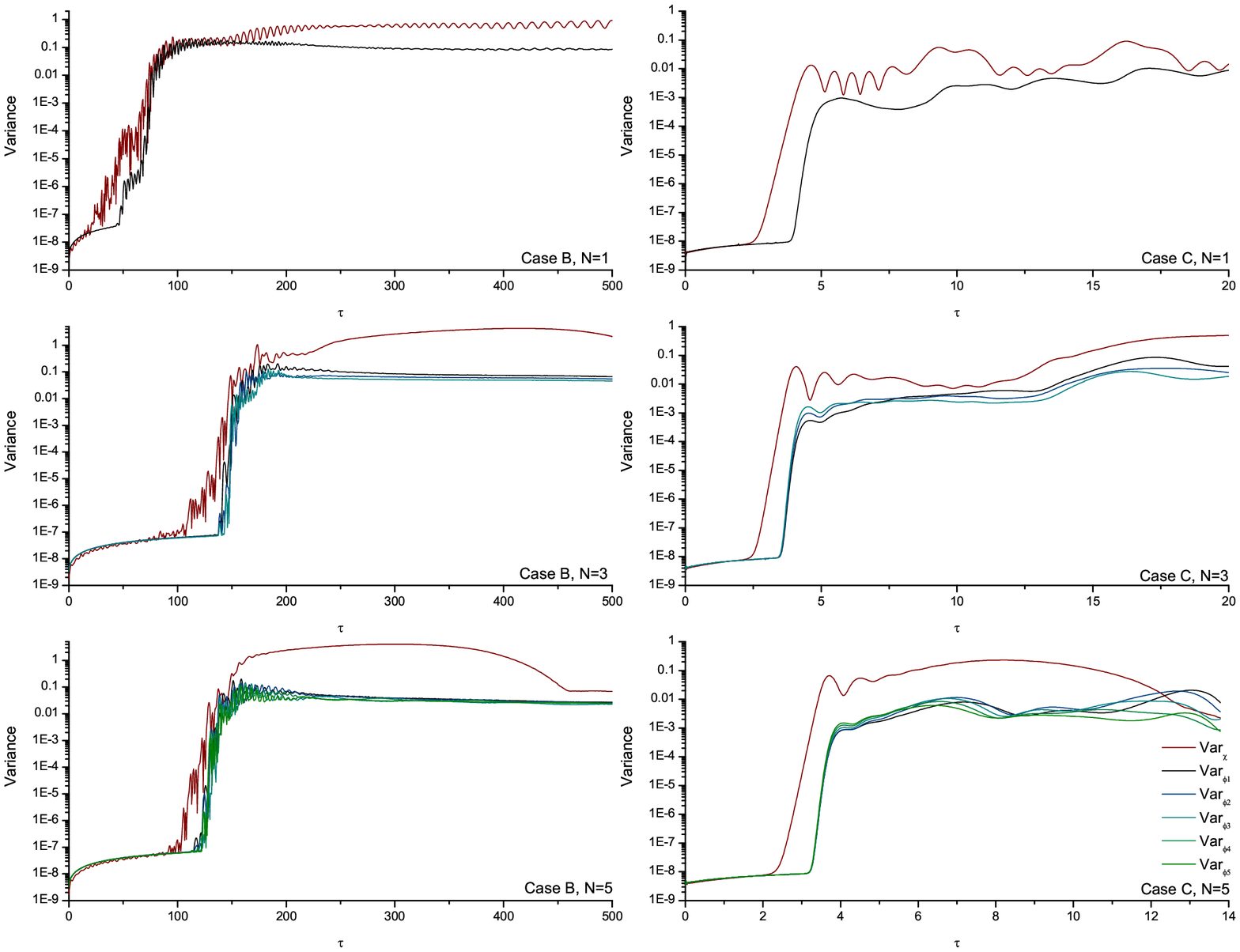}
   \caption{\label{pic:variance_B_C} The variance (\ref{variance}) of the fields for case B ($\sigma\neq 0$, $g\neq 0$, $n^3=128^3$-lattice, $L=5/m$) and case C ($g$ suppressed compared to case C), for $\mathcal{N}=1,3$ and $5$. Fields fragment and backreaction becomes important once the variance is of order one. $\sigma \neq 0$ leads to tachyonic preheating, which is efficient even in the presence of a large $g$.}
\end{figure}

In addition to the four-leg interaction, let us turn on a three-leg interaction ($\sigma\neq 0$). Consequently, the square of the effective mass picks up a contribution proportional to $\sigma\sum_i \varphi_i+g\sum_i \varphi_i^2$. The first term oscillates around zero with a damped amplitude (due to dephasing) potentially leading to tachyonic preheating, while the second one resembles a positive offset  that redshifts like matter.  

Due to the tachyonic term, we see an enhancement of preheating in Fig.~\ref{pic:energy_ratio_A_B_C_N=128} B as compared to panel A ($\sigma=0$) (compare also the variance in the left column of Fig.~\ref{pic:variance_B_C} with Fig.~\ref{pic:variance_A}); there is no discernible suppression of resonances if the number of fields is increased, but we still observe the slight delay in the onset of particle production if $\mathcal{N}$ is increased from one to two.

We conclude that incorporating a tachyonic contribution leads to efficient preheating after multi-field inflation, even in the presence of sizable four-leg interactions.

\subsection{Tachyonic Preheating ($\sigma\neq 0$, $g$ small): Slight Enhancement \label{caseC}}
Here, we investigate whether tachyonic preheating caused by a three-leg interaction ($\sigma\neq 0$) is affected by the presence of several inflaton fields if the four-leg interactions are tuned down ($g$ small)\footnote{Setting $g=0$ leads to instabilities in our code.}. Since the matter field's effective mass is dominated by a term proportional to $\sum_i \varphi_i$ instead of $\sum_i \varphi_i^2$, we do not expect any suppression: the dephasing of the fields leads to a damping of the overall amplitude of  $\sum_i \varphi_i$, which remains negative half of the time. 

As evident from Fig.~\ref{pic:energy_ratio_A_B_C_N=128} C, an instability appears fast, leading to efficient preheating within a few oscillations of the inflaton fields. The variance is plotted in Fig.~\ref{pic:variance_B_C} (right column). Interestingly, we observe a slight enhancement with an increase in $\mathcal{N}$: for a single field, a few spurts of particle production, that is more than one oscillation of the inflaton, are needed to reach $\rho_{\chi}/\rho_{\varphi}$ of order one. But, for $\mathcal{N}\geq 3$ this ratio approaches one following the first spurt of particle production and the fields fragment fast, as evident in Fig.~\ref{pic:variance_B_C}. As a consequence, the potential for the inflatons does not need to curve upwards again for $\varphi < 0$, but may simply flatten out. This enables preheating after inflation driven by fields with runaway potentials and could ease moduli trapping on the landscape \cite{Kofman:2004yc}.

\section{Discussion and Conclusion \label{sec:discussion}}
Our lattice simulations show that preheating via four-leg interactions is strongly suppressed if more than two inflatons couple to the same matter field. The cause of this suppression is a dephasing of the inflaton fields that leads to an increase of the matter field's effective mass, while smearing out oscillations in it.  One could imagine that each inflaton couples to a different preheat matter field, which in turn decay to standard model particles in the end. However, such a scenario seems fine tuned in the presence of several hundred inflatons, such as in $\mathcal{N}$-flation; even if only three inflatons, out of  many, happen to be coupled to the same matter field, they would be unable to decay during preheating, and could dominate the energy of the universe; this might be useful, since their decay, i.e. mediated by gravity, could alleviate the gravitino problem. Fields are often of the same kind in current models of multi-field inflation, for example, they are all identified with axions in $\mathcal{N}$-flation. Therefore, it seems rather unlikely for inflatons to have different decay channels.\footnote{The situation is different in models of staggered inflation \cite{Battefeld:2008py,Battefeld:2008ur,Battefeld:2008qg}, since reheating is dominated by a few long lived fields which can be quite different from the already decayed ones (i.e.~they could lie within the MSSM).}

We also considered tachyonic preheating where three-leg interactions (Yukawa couplings) lead to a tachyonic contribution to the matter field's effective mass, causing explosive particle production even if the inflatons run out of phase. In the presence of additional four-leg interactions and more fields we find no suppression. Further, if four-leg interactions are suppressed, we observe a slight enhancement in the efficiency of tachyonic preheating with increasing $\mathcal{N}$. Due to the employed tachyonic instability and the multiple fields involved, there might be additional observational signatures: the spectrum of gravitational waves produced during preheating might differ and additional non-Gaussianities might be produced. We leave these interesting avenues for future studies. 

To conclude, tachyonic preheating is a viable mechanism to transfer energy to other scalar fields after multi-field inflation, opposite to standard parametric resonance scenarios; in the latter case, we expect some (or all) fields to survive long enough for the old theory of reheating to commence. 

\begin{acknowledgments}
We thank D.~Langlois, A.~Mazumdar, and G.~Felder for discussions and S.~Kawai for comments on the draft. T.~B. is grateful for the hospitality at the Helsinki Institute of Physics and the APC (CNRS-Universit\'e Paris 7). T.~B.~is supported by the Council on Science and Technology at Princeton University. D.~B.~is supported by the EU EP6 Marie Curie Research and Training Network `UniverseNet' (MRTN-CT-2006-035863) and acknowledges the hospitality at Princeton University. JTG is supported by the Perimeter Institute for Theoretical Physics.  Research at the Perimeter Institute for Theoretical Physics is supported by the Government of Canada through Industry Canada and by the Province of Ontario through the Ministry of Research \& Innovation.

\end{acknowledgments}

\appendix
\section{Implementation of {\sc LatticeEasy}\label{appendix:code}}

The current model, omitting the trivial $\mathcal{N}=1$ case, consists of $\mathcal{N}$ inflatons, $\varphi_i$, each of which has  
\begin{equation}
m_i^2 = \frac{\mathcal{N}+2i-3}{2(\mathcal{N}-1)} m^2 \equiv \beta_i m^2\,,
\end{equation}
which describes a system that has an average square mass of 
\begin{eqnarray}
\bar{m}^2 &=&\frac{1}{\mathcal{N}}\sum_{i=1}^\mathcal{N}\beta_i m^2 \\
&=& \sum_{i=1}^\mathcal{N} \frac{(\mathcal{N}+2i-3)}{2\mathcal{N}(\mathcal{N}-1)} m^2 \\
&=& \frac{\mathcal{N}^2+\mathcal{N}(\mathcal{N}+1)-3\mathcal{N}}{2\mathcal{N}(\mathcal{N}-1)}m^2\\
&=& m^2\,,
\end{eqnarray}
as well as a lowest square mass  of $m_1^2=m^2/2$, and a greatest square mass of $m_{\mathcal{N}}^2=3m^2/2$.  The total potential energy is equally distributed among the $\mathcal{N}$ fields and sums up to $m^2 \varphi_0^2/2$,
where $\varphi_0 = 0.193$ \cite{Kofman:1997yn}, in line with single-field models.  The resulting initial potential energy
\begin{equation}
\frac{1}{2}\beta_i m^2 \varphi^2_i(0)=  \frac{1}{\mathcal{N}}\frac{1}{2} m^2 \varphi^2_0\,,
\end{equation}
yield the initial conditions
\begin{equation}
\varphi_i(0) =\sqrt{ \frac{1}{\mathcal{N}} \frac{1}{\beta_i} }\varphi_0\,,
\end{equation} 
for the zero modes of the inflaton fields.
We additionally assume that they are initially at rest, so that the homogeneous momentum vanishes, 
\begin{eqnarray}
\dot{\varphi_i}(0) = 0\,.
\end{eqnarray}
The inhomogeneous initial conditions are naturally defined in momentum space and natively calculated by {\sc LatticeEasy}.  The Fourier transform of any field in the model, $f_i$, is parameterized by an amplitude 
and a phase 
\begin{equation}
\tilde{f}_i (k)= \left|\tilde{f}_i(\vec{k})\right|e^{i\theta_i(\vec{k})}\,,
\end{equation}
 where we use the convention
\begin{equation}
f_i(\vec{x}) \equiv \frac{1}{(2\pi)^3} \int d^3k \tilde{f}_i(\vec{k}) e^{i \vec{k}\cdot\vec{x}}\,.
\end{equation}
The amplitude of each mode is chosen from a Raliegh distribution \cite{Polarski:1995jg,Khlebnikov:1996mc,Felder:2000hq},
\begin{equation}
\label{raleigh}
\mathcal{P}\left( \left|\tilde{f}_i(\vec{k})\right|\right) =  \left|\tilde{f}_i(\vec{k})\right|e^{-2a^2\omega_k \left|\tilde{f}_i(\vec{k})\right|^2}\,,
\end{equation}
where 
\begin{equation}
\omega_k = k^2 + a^2 m_{i\,{\rm eff}}^2
\end{equation}
and 
\begin{equation}
m_{i\,{\rm eff}}^2 = \frac{\partial^2 V(f_i)}{\partial f_i^2}\,.
\end{equation}
The phase $\theta_i$ is taken to be evenly distributed between $0$ and $2\pi$.  

The derivative of the field is determined analytically,
\begin{equation}
\dot{\tilde{f}}_i(k) = \pm i \omega_k \tilde{f}_i- H\tilde{f}_i\,,
\end{equation}
where the ambiguity in sign comes from choosing either a right-moving or left-moving wave.
Although  each Fourier mode of (the real field) $f_i$ {\sl should} match the Raliegh distribution, simply choosing modes from  (\ref{raleigh}) does not guarantee that the field and its derivative are real while simultaneously preserving isotropy.  Namely, the modes of the field must obey
\begin{equation}
\tilde{f}_i(-\vec{k}) = \tilde{f}_i^*(\vec{k})\,.
\end{equation}
Similarly, the modes of the derivatives must obey an analogous relation.  Felder and Tkachev impose this condition by hand, while preserving the initial spectrum of the field, by choosing two modes whose amplitudes are drawn from (\ref{raleigh}) with a random phase, $\tilde{f}_{i,1}(\vec{k})$ and $\tilde{f}_{i,2}(\vec{k})$, and defining the initial conditions of the field to be
\begin{equation}
\tilde{f}_{i}(\vec{k}) = \sqrt{\frac{1}{2}}\left(\tilde{f}_{i,1}(\vec{k}) + \tilde{f}_{i,2}(\vec{k}) \right)\,,
\end{equation}
while the initial derivatives of the fields are
\begin{equation}
\dot{\tilde{f}}_{i}(\vec{k}) = \sqrt{\frac{1}{2}}i\omega_k\left(\tilde{f}_{i,1}(\vec{k}) - \tilde{f}_{i,2}(\vec{k}) \right)-H \tilde{f}_{i}(\vec{k})\,.
\end{equation}

Lastly, we rescale time, space and fields so that the homogeneous fields and the fields' derivatives are of order one, resulting in our definition of a program time
\begin{equation}
\tau \equiv m t\,,
\end{equation}
a program comoving distance
\begin{equation}
x_{\rm pr} \equiv m x\,,
\end{equation}
and program fields 
\begin{equation}
f_{i,\rm pr} \equiv \frac{a^{3/2}}{\varphi_0} f_i\,.
\end{equation}

With the above rescaled fields, we define the average fields
\begin{eqnarray}
\bar{f}_{i,\rm pr}=\frac{1}{n^3}\sum_{j=1}^{n^3}f_{i,\rm pr}^j\,,
\end{eqnarray}
and the variance 
\begin{eqnarray}
\mbox{Var}_{f_i}=\sqrt{\frac{1}{n^3}\sum_{j=1}^{n^3}\left(\bar{f}_{i,\rm pr}-f_{i,\rm pr}^j\right)^2}\,,\label{variance}
\end{eqnarray}
where $f_{i,\rm pr}^j$ is the field value of the i'th field in the $j$'th lattice cell (we use an $n^3=128^3$ lattice so that $j=1\dots n^3$).

\end{document}